# Closures in Formal Languages and Kuratowski's Theorem

Janusz Brzozowski, Elyot Grant, and Jeffrey Shallit

David R. Cheriton School of Computer Science,
University of Waterloo, Waterloo, Ontario, Canada N2L 3G1
{brzozo, egrant, shallit}@cs.uwaterloo.ca

**Abstract.** A famous theorem of Kuratowski states that in a topological space, at most 14 distinct sets can be produced by repeatedly applying the operations of closure and complement to a given set. We re-examine this theorem in the setting of formal languages, where closure is either Kleene closure or positive closure. We classify languages according to the structure of the algebra they generate under iterations of complement and closure. We show that there are precisely 9 such algebras in the case of positive closure, and 12 in the case of Kleene closure.

## 1 Introduction

In 1922, Kuratowski proved that if $S$ is any set in a topological space, then at most 14 distinct sets can be produced by repeatedly applying the operations of topological closure and complement to $S$ [5, 2]. Furthermore, there exist sets achieving this bound of 14 in many common topological spaces.

There is a large and scattered literature on Kuratowski's theorem, most of which focuses on topological spaces; an admirable survey is the paper of Gardner and Jackson [3]. Our Theorem 1, Kuratowski's theorem in the setting of a general closure system, can be found in Hammer [4]. Our point of view most closely matches that of Peleg [6], who briefly observed that Kleene and positive closure are closure operators, and hence Kuratowski's theorem holds for them.

We reconsider this theorem in the context of formal languages, where closure is replaced by Kleene closure or positive closure. We describe all possible algebras of languages generated by a language under the operations of complement and closure. Our main results classify languages according to the structure of the algebras they generate, and give a language of each type (Theorems 5 and 7).

## 2 Closure Operators and Closure Systems

We recall the definitions and properties of closures in general. Let $S$ be a set which we call the *universal set*. An operator $\Box$ operating on a set $X \subseteq S$ will be denoted by $X^\Box$. Then a mapping $\Box : 2^S \to 2^S$ is a *closure operator* if and only if it satisfies the following for all subsets $X$ and $Y$ of $S$:

$$\begin{aligned}
&X \subseteq X^\square && (\square \text{ is extensive}); \\
&X \subseteq Y \text{ implies } X^\square \subseteq Y^\square && (\square \text{ is isotone}); \\
&X^{\square\square} = X^\square && (\square \text{ is idempotent}).
\end{aligned} \qquad (1)$$

A pair $(S, \square)$ satisfying (1) is a *closure system*. The *complement* $S \setminus X$ of a set $X \subseteq S$ is denoted $X^-$. The set $X^\square$ is the *closure* of $X$. We say $X$ is *closed* if $X = X^\square$. Also, $X$ is *open* if its complement is closed, and $X$ is *clopen* if it is both open and closed. The *interior* of $X$ (denoted $X^\circ$) is defined to be $X^{-\square-}$.

Note the duality between $\square$ and $\circ$:

$$X^\circ = X^{-\square-} \qquad X^\square = X^{-\circ-}.$$

This duality also applies to (1), since we have

$$\begin{aligned}
&X \supseteq X^\circ && (\circ \text{ is intensive}); \\
&X \subseteq Y \text{ implies } X^\circ \subseteq Y^\circ && (\circ \text{ is isotone}); \\
&X^{\circ\circ} = X^\circ && (\circ \text{ is idempotent}).
\end{aligned} \qquad (2)$$

Moreover, it is equivalent to define $(S, \square)$ via an *interior operator* satisfying (2).

We now list some fundamental properties of closure systems; note the duality.

**Proposition 1.** *(a) The intersection of an arbitrary family of closed sets is closed. (b) The union of an arbitrary family of open sets is open.*

**Proposition 2.** *For $X \subseteq S$, the following are identical: (a) $X^\square$; (b) $\bigcap \{Y \subseteq S : Y \supseteq X \text{ and } Y \text{ is closed}\}$; (c) $\{a \in S : \text{for all open } Y \subseteq S, a \in Y \text{ implies } Y \cap X \neq \emptyset\}$; (d) $X^{-\circ-}$.*

**Proposition 3.** *For $X \subseteq S$, the following are identical: (a) $X^\circ$; (b) $\bigcup \{Y \subseteq S : Y \subseteq X \text{ and } Y \text{ is open}\}$; (c) $\{a \in S : \text{there exists an open } Y \subseteq X, \text{with } a \in Y\}$; (d) $X^{-\square-}$.*

**Proposition 4.** *Let $X, Y \subseteq S$. Then the following hold:*

*(a) $X^\square$ is closed.*
*(b) $(X \cup Y)^\square = (X^\square \cup Y^\square)^\square$.*
*(c) $(X \cap Y)^\square \subseteq X^\square \cap Y^\square$.*

**Proposition 5.** *Let $X, Y \subseteq S$. Then the following hold:*

*(a) $X^\circ$ is open.*
*(b) $(X \cap Y)^\circ = (X^\circ \cap Y^\circ)^\circ$.*
*(c) $(X \cup Y)^\circ \supseteq X^\circ \cup Y^\circ$.*

A closure operator is *topological* if the union of a finite family of closed sets is closed (or, equivalently, if the intersection of a finite family of open sets is open). The closure operators which we discuss in the context of formal languages are *not* topological, and so we will not assume this property holds.

## 3   Kuratowski's Complement-Closure Theorem

Here, we state two versions of Kuratowski's theorem. The first [5] is equivalent to his original result generalized to an arbitrary closure system, not necessarily topological:

**Theorem 1.** *Let $(S, \Box)$ be a closure system, and let $X \subseteq S$. Starting with $X$, apply the operations of closure and complement in any order, any number of times. Then at most* 14 *distinct sets are generated. Also, any $X \subseteq S$ satisfies:*

$$X^{\Box-\Box-\Box-\Box} = X^{\Box-\Box}. \tag{3}$$

A closure operator $\Box$ *preserves openness* if $X^\Box$ is open for all open sets $X$, or equivalently, if $Y^\circ$ is closed for all closed sets $Y$. Hence if $\Box$ preserves openness, then $X^{\Box\circ}$ and $X^{\circ\Box}$ are clopen for all sets $X$. We will see later that positive closure of languages preserves openness.

**Theorem 2.** *Let $(S, \Box)$ be a closure system such that $\Box$ preserves openness, and let $X \subseteq S$. Starting with $X$, apply the operations of closure and complement in any order, any number of times. Then at most* 10 *distinct sets are generated. Also, any $X \subseteq S$ satisfies:*

$$X^{\Box-\Box-\Box} = X^{\Box-\Box-}. \tag{4}$$

*Proof.* Let $\mathcal{F} = \{X, X^\Box, X^\circ, X^{\circ\Box}, X^{\Box\circ}, X^-, X^{\Box-}, X^{\circ-}, X^{\circ\Box-}, X^{\Box\circ-}\}$ be a family of sets, not necessarily all distinct. Clearly, $\mathcal{F}$ is closed under complement. If $\Box$ preserves openness, then $X^{\circ\Box}$ and $X^{\Box\circ}$ are both clopen. Hence $X^{\circ\Box}$, $X^{\Box\circ}$, $X^{\circ\Box-}$, and $X^{\Box\circ-}$ are all closed and thus equal to their closures. Equation (4) then follows since $X^{\Box-\Box-} = X^{\Box\circ}$. Moreover, $X^\Box$ and $X^{\circ-}$ are also closed, leaving $X^-$ and $X^{\Box-}$ as the only sets whose closures could possibly lie outside $\mathcal{F}$. But $X^{-\Box} = X^{\circ-}$ and $X^{\Box-\Box} = X^{\Box\circ-}$, so $\mathcal{F}$ must be closed under both complement and closure. Hence only the sets in $\mathcal{F}$ may be generated from $X$ by repeatedly applying complement and closure, and the result follows. ∎

In 1983, Peleg [6] defined a closure operator to be *compact* if it satisfies Equation 4. He showed that at most 10 different sets are generated if $\Box$ is compact, and proved that $\Box$ preserves openness if and only if it is compact. Thus Theorem 2 is a modified version of Peleg's result.

## 4   Positive and Kleene Closures of Languages

We deal now with closures in the setting of formal languages. Our universal set is $\Sigma^*$, the set of all finite words over a finite non-empty alphabet $\Sigma$. We consider two closure operators: positive closure and Kleene closure. For $L \subseteq \Sigma^*$, we define $L^- = \Sigma^* \setminus L$, $L^+ = \bigcup_{i \geq 1} L^i$, and $L^* = \bigcup_{i \geq 0} L^i$.

**Proposition 6.** *Let $L, M \subseteq \Sigma^*$. Then the following hold:*

(a) $L \subseteq L^+$ and $L \subseteq L^*$.
(b) $L \subseteq M$ implies $L^+ \subseteq M^+$ and $L^* \subseteq M^*$.
(c) $L^{++} = L^+$ and $L^{**} = L^*$.

**Corollary 1.** *Positive closure and Kleene closure are both closure operators.*

We note, importantly, that the positive and Kleene closures are *not* topological. As a counterexample, observe that $(aa)^+ \cup (aaa)^+ \subsetneq (aa \cup aaa)^+$, as $a^5$ belongs to the right-hand side but not the left. Consequently, languages *do not* form a topology under positive or Kleene closure.

A language is *positive-closed* if it is a closed set under positive closure. We define *positive-open*, *Kleene-closed*, and *Kleene-open* similarly. The *positive interior* of a language $L$ is $L^\oplus = L^{-+-}$; the *Kleene interior* is $L^\circledast = L^{-*-}$.

**Proposition 7.** *Let $L \subseteq \Sigma^*$. The following are equivalent:*

(a) *$L$ is positive-closed.*
(b) *$L \cup \{\epsilon\}$ is Kleene-closed.*
(c) *$L = L^+$.*
(d) *$L = M^+$ for some $M \subseteq \Sigma^*$.*
(e) *For all $u, v \in L$, we have $uv \in L$.*

**Proposition 8.** *Let $L \subseteq \Sigma^*$. The following are equivalent:*

(a) *$L$ is positive-open.*
(b) *$L \setminus \{\epsilon\}$ is Kleene-open.*
(c) *$L = L^\oplus$.*
(d) *$L = M^\oplus$ for some $M \subseteq \Sigma^*$.*
(e) *For all $u, v \in \Sigma^*$ such that $uv \in L$, we have $u \in L$ or $v \in L$.*

In algebraic terms, a language $L \subseteq \Sigma^*$ is a *semigroup* if $uv \in L$ for all $u, v \in L$. Proposition 7 therefore states that a language is positive-closed if and only if it is a semigroup. Additionally, $L \subseteq \Sigma^*$ is Kleene-closed if and only if it is a *monoid*—a semigroup containing $\epsilon$. We easily verify that if $L$ is positive-closed, then so are $L \setminus \{\epsilon\}$ and $L \cup \{\epsilon\}$. Thus there is an obvious 2-to-1 mapping between positive-closed and Kleene-closed languages—positive-closed languages may or may not contain $\epsilon$, and Kleene-closed languages must contain $\epsilon$.

Since positive closure and Kleene closure are so similar, we shall restrict our attention to positive closure from this point onward. This will allow us to state our theorems more elegantly, as we will not need to worry about $\epsilon$. For the remainder of this article, a language is *closed* if it is positive-closed, *open* if it is positive-open, and *clopen* if it is both positive-closed and positive-open.

*Example 1.* We first give an example of a clopen language. Let $\Sigma$ be an alphabet and let $\Sigma_1, \Sigma_2 \subseteq \Sigma$. For $w \in \Sigma^*$, let $|w|_1$ (respectively, $|w|_2$) denote the number of distinct values of $i$ for which $w[i] \in \Sigma_1$ (respectively, $w[i] \in \Sigma_2$). Suppose $k \geq 0$. Then $\{w \in \Sigma^* : |w|_1 < k|w|_2\}$ is clopen.

To prove this, we let $L = \{w \in \Sigma^* : |w|_1 < k|w|_2\}$. Let $u, v \in L$. Then $|u|_1 < k|u|_2$ and $|v|_1 < k|v|_2$. But $|uv|_1 = |u|_1 + |v|_1 < k|u|_2 + k|v|_2 = k|uv|_2$, so $uv \in L$, and thus $L$ is closed. By a similar argument, we can prove that $L^- = \{w \in \Sigma^* : |w|_1 \geq k|w|_2\}$ is closed. Thus $L$ is clopen. ∎

*Example 2.* Next, we give several examples of open languages. A language $L$ is *prefix-closed* if and only if for every $w \in L$, each non-empty prefix of $w$ is in $L$. We analogously define *suffix-closed*, *subword-closed*, and *factor-closed* languages. Here by subword, we mean an arbitrary subsequence, and by factor, we mean a contiguous subsequence. For any $L \subseteq \Sigma^*$, if $L$ is prefix-, suffix-, factor-, or subword-closed, then $L$ is open.

To verify the claim for prefix-closed languages, we show that $L$ satisfies Proposition 8 (e). Let $w \in L$ and suppose $w = uv$. Then $u \in L$ if $L$ is prefix-closed, so our characterization holds and $L$ is open. The proof is similar if $L$ is suffix-closed. Since factor- and subword–closed languages are also prefix-closed, the claim holds. ∎

*Example 3.* Languages that are left ideals (have the form $L = \Sigma^* L$), right ideals ($L = L\Sigma^*$), two-sided ideals ($L = \Sigma^* L \Sigma^*$), or have the form $L = \bigcup_{a_1 \cdots a_n \in L} \Sigma^* a_1 \Sigma^* \cdots \Sigma^* a_n \Sigma^*$, all satisfy $L = L^+$, and so are positive closed. ∎

In the 1970's, D. Forkes proved Equation 3 with the Kleene closure as $\square$, and the first author then proved that Equation 4 holds when $\square$ is positive closure. (They were both unaware of [5].) Peleg [6] proved this in greater generality over a wider class of operators. Here, we give a proof of an equivalent fact: positive closure preserves openness.

**Theorem 3.** *Let $L \subseteq \Sigma^*$ be open. Then $L^+$ is open.*

*Proof.* To show this, we use our characterization of open languages given in Proposition 8 (e). Let $u, v \in \Sigma^*$ be such that $uv \in L^+$. Then let $uv = x_1 x_2 \cdots x_n$ where each $x_i \in L$. There must exist an index $j$, $1 \leq j \leq n$, such that $u = x_1 \cdots x_{j-1} x'_j$, $v = x''_j x_{j+1} \cdots x_n$, and $x'_j x''_j = x_j$. But since $x'_j x''_j = x_j \in L$ and $L$ is open, either $x'_j \in L$ or $x''_j \in L$. Thus either $u = x_1 \cdots x_{j-1} x'_j \in L^+$ or $v = x''_j x_{j+1} \cdots x_n \in L^+$. Hence $L^+$ is open by our characterization. ∎

By arguments similar to those in the proof of Theorem 2, we may conclude:

**Corollary 2.** *Let $L \subseteq \Sigma^*$. Then $L^{+\oplus}$ and $L^{\oplus+}$ are clopen. Moreover, if $L$ is open, then $L^+$ is clopen, and if $L$ is closed, then $L^\oplus$ is clopen.*

The converses of the above results are false; for example, there exist languages such as $\{a, aaaa\}$ which are not open but have clopen closures. We will discuss such possibilities extensively in the next section. For now, we give a characterization of the languages with clopen closures and clopen interiors.

**Theorem 4.** *Let $L \subseteq \Sigma^*$.*

*(a) $L^+$ is clopen iff there exists an open language $M$ with $L \subseteq M \subseteq L^+$.*

(b) $L^\oplus$ is clopen iff there exists a closed language $M$ with $L \supseteq M \supseteq L^\oplus$.

*Proof.* We prove only (a); (b) can be proved using a similar argument. The forward direction of (a) is trivial since we can take $M = L^+$. For the converse, we note that $L \subseteq M$ implies $L^+ \subseteq M^+$ by isotonicity, and $M \subseteq L^+$ implies $M^+ \subseteq L^{++} = L^+$ by isotonicity and idempotency. Thus $M^+ = L^+$, and since $M^+$ is the closure of an open language, it is clopen and the result follows. ∎

## 5 Kuratowski's Theorem for Languages

For any language $L$, let $A(L)$ be the family of all languages generated from $L$ by complementation and positive closure. Since positive closure preserves openness, Theorem 2 implies that $A(L)$ contains at most 10 languages. As we shall see, this upper bound is tight. Moreover, there are precisely 9 distinct finite algebras $(A(L),^+,^-)$. Since the languages in $A(L)$ must occur in complementary pairs, there can only exist algebras containing 2, 4, 6, 8, or 10 distinct languages. We will provide a list of conditions that classify languages according to the structure of $(A(L),^+,^-)$, and thus completely describe the circumstances under which $|A(L)|$ is equal to 2, 4, 6, 8, or 10.

We will also explore Kleene closure, where there are subtle differences. Let $D(L)$ be the family of all languages generated from $L$ by complementation and Kleene closure. Kleene closure does not preserve openness, since Kleene-closed languages contain $\epsilon$ and Kleene-open languages do not. Therefore we must fall back to Theorem 1, which implies that $D(L)$ contains at most 14 languages, and we will show that this bound is also tight. There are precisely 12 distinct finite algebras $(D(L),^*,^-)$. We shall describe these algebras by relating them to those in the positive case.

In a sense, our results are the formal language analogue of topological results obtained by Chagrov [1] and discussed in [3]. Peleg [6] noted the tightness of the bounds of 10 and 14 in the positive and Kleene cases, but went no further.

### 5.1 Structures of the Algebras with Positive Closure

We may better understand the structure of $A(L)$ by first analyzing a related algebra of languages. Let $B(L)$ be the family of all languages generated from $L$ by positive closure and positive interior, and let $C(L) = \{M : M^- \in B(L)\}$ be their complements. Recall that the closure of an open language is clopen and the interior of a closed language is clopen by Corollary 2. Since the closure and interior operators are idempotent on the clopen languages $L^{+\oplus}$ and $L^{\oplus+}$, it follows that $B(L) = \{L, L^+, L^{+\oplus}, L^\oplus, L^{\oplus+}\}$. Of course, these five languages may not all be distinct; we will address this later. At the moment, we provide the following proposition, which demonstrates that it suffices to analyze the structure of $B(L)$ to determine the structure of $A(L)$.

**Proposition 9.** *Let $L \subseteq \Sigma^*$. Then $A(L) = B(L) \cup C(L)$, and the union is disjoint.*

*Proof.* Clearly $A(L) \supseteq B(L) \cup C(L)$, since any language generated from $L$ by closure, interior, and complement can be generated using only closure and complement by the identity $L^\oplus = L^{-+-}$. To prove the reverse inclusion, we let $M \in A(L)$. Then there is some string of symbols $z \in \{+, -\}^*$ such that $M = L^z$. We construct a string $z' \in \{+, -, \oplus\}^*$ by starting with $z$ and repeatedly replacing all instances of $-+$ by $\oplus-$ and all instances of $-\oplus$ by $+-$, until no such replacements are possible. Since $L^{-+} = L^{\oplus-}$ and $L^{-\oplus} = L^{+-}$, we have $M = L^{z'}$. However, in producing $z'$, we effectively shuffle all complements to the right. Consequently, the operation performed by $z'$ is a series of positive closures and interiors followed by an even or odd number of complements. Hence either $M \in B(L)$ or $M \in C(L)$, and thus $A(L) = B(L) \cup C(L)$.

We now prove that $B(L) \cap C(L) = \emptyset$. We assume otherwise to obtain a contradiction; $B(L)$ must then contain some complementary pair of languages $M$ and $M^-$. We note that $L^\oplus \subseteq L^{\oplus+}$ by extensivity, $L^\oplus \subseteq L \subseteq L^+$ by intensivity and extensivity, and $L^\oplus \subseteq L^{+\oplus}$ by isotonicity, and hence $L^\oplus \subseteq M$ for all $M \in B(L)$. Thus for two languages in $B(L)$ to be complements, $L^\oplus$ must be empty. Then $L$ contains no strings of length 1, and hence $L^+$ and $L^{+\oplus}$ do not either. But then no language in $B(L)$ contains a string of length 1, and thus no pair of languages in $B(L)$ are complements, and we have our contradiction. ∎

Proposition 9 implies that $|A(L)| = 2|B(L)|$, and moreover that there is an exact 1-to-2 correspondence between the languages in $B(L)$ and $A(L)$: each language in $B(L)$ can be associated with itself and its complement. Hence the algebra $(A(L), {}^+, {}^-)$ can be constructed by simply merging the two algebras $(B(L), {}^+, {}^\oplus)$ and $(C(L), {}^+, {}^\oplus)$ and adding the complement operator. Thus we have reduced the problem of describing all algebras $(A(L), {}^+, {}^-)$ to the simpler task of describing the algebras $(B(L), {}^+, {}^\oplus)$. Before we proceed, we need to exclude a possible case via the following:

**Lemma 1.** *Suppose $L \subseteq \Sigma^*$. If $L^+$ and $L^\oplus$ are both clopen, then $L$ must be open or closed.*

*Proof.* Seeking a contradiction, we assume that both $L^+$ and $L^\oplus$ are clopen but $L$ is neither open nor closed. If $L$ is not open, then $L \setminus L^\oplus$ is non-empty.

Let $w$ be the shortest word in $L \setminus L^\oplus$. Consider $M = L^\oplus \cup \{w\}$. It must not be open, because if it were, we would have $M \subseteq L^\oplus$ by Proposition 3. Then Proposition 8 (e) must fail to hold for some word in $M$. But it holds for all words in $L^\oplus$ and thus must fail for $w$. Then there exists nonempty words $x$ and $y$ with $xy = w$, but $x \notin M$ and $y \notin M$. Then neither $x$ nor $y$ is in $L^\oplus$.

By our assumption that $L^+$ is open, the fact that $w \in L^+$ implies that either $x \in L^+$ or $y \in L^+$. Suppose without loss of generality that $x \in L^+$. Then $x$ is the concatenation of a list of words from $L$; we write $x = u_1 u_2 \cdots u_n$ with $u_i \in L$ for all $1 \leq i \leq n$. Then $|u_i| \leq |x| < |w|$ for all $i$ and thus $u_i \in L^\oplus$ for all $i$ by our definition of $w$ as the shortest word in $L \setminus L^\oplus$. However, $x$ is then the concatenation of a list of words from $L^\oplus$ and is thus an element of $L^{\oplus+}$, which is $L^\oplus$ since we assumed $L^\oplus$ was closed. This is a contradiction since $x \notin L^\oplus$. ∎

Finally, we characterize the 9 possible algebras $(B(L), ^+, ^\oplus)$. Table 1 classifies all languages according to the structures of the algebras they generate and gives an example of each type. Here, we briefly explain our analysis. Clearly $B(L) = \{L\}$ if and only if $L$ is clopen, giving case (1). If $L$ is open but not closed, then $B(L) = \{L, L^+\}$ since $L^+$ must then be clopen. Similarly, if $L$ is closed but not open, then $B(L) = \{L, L^\oplus\}$. These situations yield cases (2) and (3). We henceforth assume that $L$ is neither open nor closed, and thus $L$, $L^\oplus$, and $L^+$ are all different. The remaining cases depend on the values of $L^{\oplus+}$ and $L^{+\oplus}$. Both must be clopen, so neither can equal $L$. Lemma 1 proves that $L^\oplus$ and $L^+$ cannot both be clopen. If neither $L^\oplus$ nor $L^+$ are clopen, then we have case (8) if $L^{\oplus+}$ and $L^{+\oplus}$ are equal and case (9) if they are not. The remaining cases occur when one of $L^+$ and $L^\oplus$ is clopen and the other is not. If $L^+$ is clopen and $L^\oplus$ is not, then we get case (4) if $L^{\oplus+} = L^+$ and case (6) otherwise. Analogously, if $L^\oplus$ is clopen and $L^+$ is not, then we get case (5) if $L^{+\oplus} = L^\oplus$ and case (7) otherwise.

| Case | Necessary and Sufficient Conditions | $|B(L)|$ | $|A(L)|$ | Example | Dual |
|---|---|---|---|---|---|
| (1) | $L$ is clopen. | 1 | 2 | $a^*$ | (1) |
| (2) | $L$ is open but not closed. | 2 | 4 | $a$ | (3) |
| (3) | $L$ is closed but not open. | 2 | 4 | $aaa^*$ | (2) |
| (4) | $L$ is neither open nor closed; $L^+$ is clopen and $L^{\oplus+} = L^+$. | 3 | 6 | $a \cup aaa$ | (5) |
| (5) | $L$ is neither open nor closed; $L^\oplus$ is clopen and $L^{+\oplus} = L^\oplus$. | 3 | 6 | $aa$ | (4) |
| (6) | $L$ is neither open nor closed; $L^+$ is open but $L^\oplus$ is not closed; $L^{\oplus+} \neq L^+$. | 4 | 8 | $a \cup abaa$ | (7) |
| (7) | $L$ is neither open nor closed; $L^\oplus$ is closed but $L^+$ is not open; $L^{+\oplus} \neq L^\oplus$. | 4 | 8 | $(a \cup b)^* \setminus (a \cup abaa)$ | (6) |
| (8) | $L$ is neither open nor closed; $L^\oplus$ is not closed and $L^+$ is not open; $L^{+\oplus} = L^{\oplus+}$. | 4 | 8 | $a \cup bb$ | (8) |
| (9) | $L$ is neither open nor closed; $L^\oplus$ is not closed and $L^+$ is not open; $L^{+\oplus} \neq L^{\oplus+}$. | 5 | 10 | $a \cup ab \cup bb$ | (9) |

**Table 1.** Classification of languages by the structure of $(B(L), ^+, ^\oplus)$

We see that if $(B(L), ^+, ^\oplus)$ has algebraic structure (2), then $(C(L), ^+, ^\oplus)$ has structure (3). Thus we shall say that case (3) is the *dual* of case (2). By examining the conditions under which each case holds, we can easily see that cases (4) and (5) are also duals, as are cases (6) and (7). Cases (1), (8), and (9) are self-dual. This notion is useful in constructing the algebra $(A(L), ^+, ^-)$; we connect an instance of $(B(L),^+,^\oplus)$ to its dual structure in the obvious way via the complement operator. Figure 1 gives an example of this for case (6).

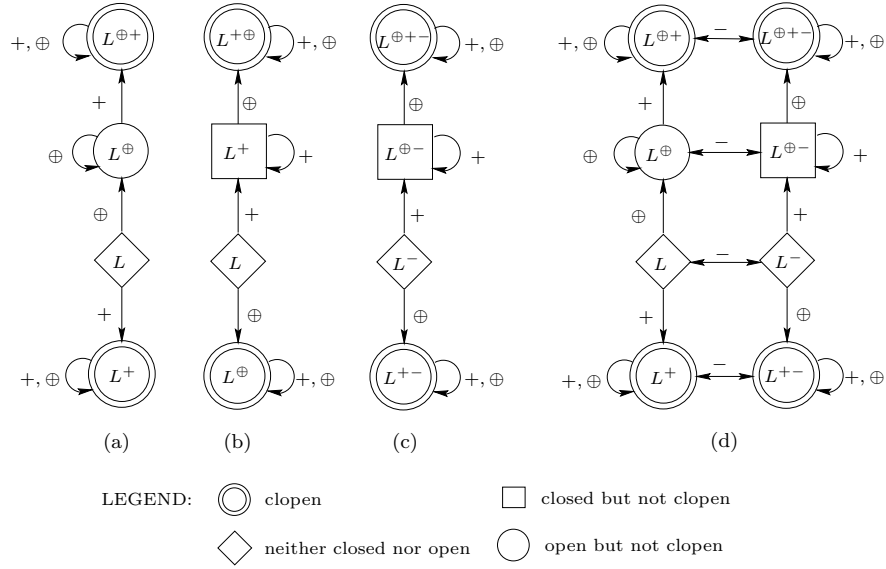

**Fig. 1.** Construction of $A(L)$, Case (6): (a) $(B(L),^+,^\oplus)$, Case (6); (b) $(B(L),^+,^\oplus)$, Case (7), the dual of Case (6) obtained by interchanging $+$ with $\oplus$, and "open" with "closed"; (c) $(C(L),^+,^\oplus)$, that is, $(B(L),^+,^\oplus)$, Case (7), with elements renamed as complements of those of Case (6); (d) $A(L)$ constructed from $B(L)$ and $C(L)$.

In summary, we have proven

**Theorem 5.** *Start with any language $L$, and apply the operators of positive closure and complement in any order, any number of times. Then at most 10 distinct languages are generated, and this bound is optimal. Furthermore, Table 1 classifies languages according to the algebra they generate and gives a language generating each algebra.*

In the unary case, we obtain the following:

**Theorem 6.** *Start with any unary language $L$, and apply the operators of positive closure and complement in any order, any number of times. Then at most 6 distinct languages are generated, and this bound is optimal. Furthermore, precisely cases (1) through (5) in Table 1 are possible for a unary language.*

*Proof.* We assume that $L \subseteq a^*$ and consider the following two possibilities:

Case (i): $a \in L$. Then $L^+ = aa^*$ or $L^+ = a^*$, both of which are clopen. Furthermore, $a \in L^\oplus$ and thus $L^{\oplus+} = L^+$. Hence one of cases (1), (2), or (4) must hold and so $|A(L)| \leq 6$.

Case (ii): $a \notin L$. Then $L^\oplus = \emptyset$ or $L^\oplus = \{\epsilon\}$, both of which are clopen. Furthermore, $a \notin L^+$ and thus $L^{+\oplus} = L^+$. Hence one of cases (1), (3), or (5) must hold and so $|A(L)| \leq 6$.

Unary examples for cases (1) through (5) can be found in Table 1. ∎

## 5.2 Structures of the Algebras with Kleene Closure

As we did in the positive case, first we restrict ourselves to closure and interior. Let $E(L)$ be the family of all languages generated from $L$ by Kleene closure and Kleene interior, and let $F(L) = \{M : M^- \in E(L)\}$ be their complements. Our next results relate $D(L)$ and $E(L)$ to $A(L)$ and $B(L)$. Our discussion involves both closure operators, so we will be explicit about which closure properties we are invoking (although the word *clopen* will still mean positive-clopen). We first claim the following, which can be proven in the same manner as Proposition 9:

**Proposition 10.** *Let $L \subseteq \Sigma^*$. Then $D(L) = E(L) \cup F(L)$, and the union is disjoint.*

Next, we give a way of relating $E(L)$ to $B(L)$. We recall that $L^* = L^+ \cup \{\epsilon\}$ and $L^\circledast = L^\oplus \setminus \{\epsilon\}$. Consequently, $E(L) \subseteq \bigcup_{M \in B(L)} \{M \cup \{\epsilon\}, M \setminus \{\epsilon\}\}$. We now know enough to explicitly determine $D(L)$ in the following case:

**Proposition 11.** *Let $L \subseteq \Sigma^*$ be clopen. Then $D(L) = \{L \cup \{\epsilon\}, L \setminus \{\epsilon\}, L^- \cup \{\epsilon\}, L^- \setminus \{\epsilon\}\}$.*

*Proof.* If $L$ is clopen, then $L = L^+$ and $L = L^\oplus$. Thus $L^* = L \cup \{\epsilon\}$ and $L^\circledast = L \setminus \{\epsilon\}$. These values thus form $E(L)$ and their complements form $F(L)$. Our result then follows from Proposition 10. ∎

Since the operations of positive closure and positive interior preserve the presence or absence of $\epsilon$ in a language, we may also note that if $\epsilon \in L$, then all languages in $B(L)$ contain $\epsilon$, and conversely if $\epsilon \notin L$, then no language in $B(L)$ contains $\epsilon$. For $M \in E(L)$, we write $\phi(M)$ to denote either $M \cup \{\epsilon\}$ or $M \setminus \{\epsilon\}$, whichever lies in $B(L)$. We note that $\phi(M)$, $\phi(M \cup \{\epsilon\})$, and $\phi(M \setminus \{\epsilon\})$ are equal. Moreover, we note that $\phi(M^*) = \phi(M)^+$ and $\phi(M^\circledast) = \phi(M)^\oplus$; $\phi$ can therefore be thought of as a homomorphism from $E(L)$ to $B(L)$. Consequently, $E(L) \subseteq \{M : \phi(M) \in B(L)\}$. We use this idea and the classifications of Table 1 to determine all possible algebras $(E(L), ^*, ^\circledast)$. As we shall see, there are precisely 12 distinct algebras, each containing at most 14 elements.

We have seen what happens in case (1) when $L$ is clopen; 2 algebras are possible depending on whether $\epsilon \in L$ or not, and we refer to these as cases (1a) and (1b) respectively. We next examine cases (2) and (3), in which $L$ is not clopen but is open or closed. Suppose $L$ is open but not clopen, and hence $B(L) = \{L, L^+\}$. Then $L^*$ is clopen and thus $E(L^*) = \{L^*, L^* \setminus \{\epsilon\}\}$. Since $E(L^*) \subseteq E(L)$ we thus have $\{L, L^*, L^* \setminus \{\epsilon\}\} \subseteq E(L) \subseteq \{M : \phi(M) \in \{L, L^+\}\}$. Therefore, we have two cases; either one or both of $L \setminus \{\epsilon\}$ and $L \cup \{\epsilon\}$ may be in $E(L)$, depending on whether or not $L^\circledast = L$. If $\epsilon \notin L$, then $L^\circledast = L$ and thus $E(L) = \{L, L^*, L^* \setminus \{\epsilon\}\}$. If $\epsilon \in L$, then $L^\circledast = L \setminus \{\epsilon\}$ and thus $E(L) = \{L, L \setminus \{\epsilon\}, L^*, L^* \setminus \{\epsilon\}\}$. We refer to these situations as cases (2a) and (2b) respectively.

Similar possibilities occurs when $L$ is closed but not clopen. If $\epsilon \in L$ then $E(L) = \{L, L^\circledast, L^\circledast \cup \{\epsilon\}\}$. If $\epsilon \notin L$ then $L^* = L \cup \{\epsilon\}$ and thus $E(L) = \{L, L \cup \{\epsilon\}, L^\circledast L^\circledast \cup \{\epsilon\}\}$. We refer to these situations as cases (3a) and (3b) respectively.

We now turn our attention to cases (4) through (9), when $L$ is neither closed nor open.

**Lemma 2.** *Let $L \subseteq \Sigma^*$ be neither open nor closed. Then*
$$E(L) = \{L\} \cup \{M \cup \{\epsilon\} : M \in B(L) \text{ and } M \text{ closed}\}$$
$$\cup \{M \setminus \{\epsilon\} : M \in B(L) \text{ and } M \text{ open}\}.$$

*Proof.* Clearly $L \in E(L)$. We claim that no other language $M$ with $\phi(M) = L$ can be in $E(L)$. If we suppose otherwise, then such an $M$ must be generated by taking the Kleene closure or interior of some other language in $E(L)$. This would imply that $M$ is open or closed, which is impossible since $\phi(M) = L$ and $L$ is neither open nor closed.

For each remaining $M \in B(L) \setminus \{L\}$, we wish to show that $M \cup \{\epsilon\} \in E(L)$ if and only if $M$ is closed, and $M \setminus \{\epsilon\} \in E(L)$ if and only if $M$ is open. Let $M \in B(L) \setminus \{L\}$ be generated by some non-empty sequence $S$ of positive closures and positive interiors. If we replace each positive closure by a Kleene closure and each positive interior by a Kleene interior, then we will obtain a sequence $S'$ that generates some $M' \in E(L)$ with $\phi(M') = M$. Now $M'$ contains $\epsilon$ if and only if the last operation in $S'$ was a Kleene closure. If $M$ is closed, we may append a final positive closure to any such $S$ to obtain one in which the last operation is a closure. Conversely, if there exists an $S$ whose last operation is a closure, then $M$ must be closed. Thus there exists an $M' \in E(L)$ containing $\epsilon$ with $\phi(M') = M$ if and only if $M$ is closed. By a similar argument, there exists an $M' \in E(L)$ not containing $\epsilon$ with $\phi(M') = M$ if and only if $M$ is open. The result follows. ∎

Lemma 2 allows us to describe the structure of the algebra $(E(L), ^*, ^\circledast)$ in cases (4) through (9). Algebra $E(L)$ contains $M \cup \{\epsilon\}$ for all closed $M$ in $B(L)$, $M \setminus \{\epsilon\}$ for all open $M$ in $B(L)$, and both for all clopen $M$ in $B(L)$.

We summarize the 12 distinct algebras in Table 2. The conditions are identical to those found in Table 1; the only differences lie in cases (1), (2), and (3), where the initial presence or absence of $\epsilon$ can affect the structure of the algebra.

We now summarize our results for the Kleene case:

**Theorem 7.** *Start with any language $L$, and apply the operators of Kleene closure and complement in any order, any number of times. Then at most 14 distinct languages are generated, and this bound is optimal. Furthermore, Table 2 describes the 12 algebras generated by this process, classifies languages according to the algebra they generate, and gives a language generating each algebra.*

**Acknowledgments:** This research was supported by the Natural Sciences and Engineering Research Council of Canada.

| Case | Necessary and Sufficient Conditions | $|E(L)|$ | $|D(L)|$ | Example | Dual |
|---|---|---|---|---|---|
| (1a) | $L$ is clopen; $\epsilon \in L$. | 2 | 4 | $a^*$ | (1b) |
| (1b) | $L$ is clopen; $\epsilon \notin L$. | 2 | 4 | $a^+$ | (1a) |
| (2a) | $L$ is open but not clopen; $\epsilon \in L$. | 3 | 6 | $a \cup \epsilon$ | (3a) |
| (2b) | $L$ is open but not clopen; $\epsilon \notin L$. | 4 | 8 | $a$ | (3b) |
| (3a) | $L$ is closed but not clopen; $\epsilon \notin L$. | 3 | 6 | $aaa^*$ | (2a) |
| (3b) | $L$ is closed but not clopen; $\epsilon \in L$. | 4 | 8 | $aaa^* \cup \epsilon$ | (2b) |
| (4) | $L$ is neither open nor closed; $L^+$ is clopen and $L^{\oplus +} = L^+$. | 4 | 8 | $a \cup aaa$ | (5) |
| (5) | $L$ is neither open nor closed; $L^\oplus$ is clopen and $L^{+\oplus} = L^\oplus$. | 4 | 8 | $aa$ | (4) |
| (6) | $L$ is neither open nor closed; $L^+$ is open but $L^\oplus$ is not closed; $L^{\oplus +} \neq L^+$. | 6 | 12 | $a \cup abaa$ | (7) |
| (7) | $L$ is neither open nor closed; $L^\oplus$ is closed but $L^+$ is not open; $L^{+\oplus} \neq L^\oplus$. | 6 | 12 | $(a \cup b)^* \setminus (a \cup abaa)$ | (6) |
| (8) | $L$ is neither open nor closed; $L^\oplus$ is not closed and $L^+$ is not open; $L^{+\oplus} = L^{\oplus +}$. | 5 | 10 | $a \cup bb$ | (8) |
| (9) | $L$ is neither open nor closed; $L^\oplus$ is not closed and $L^+$ is not open; $L^{+\oplus} \neq L^{\oplus +}$. | 7 | 14 | $a \cup ab \cup bb$ | (9) |

**Table 2.** Classification of languages by the structure of $(E(L), ^*, ^\circledast)$